\documentclass[conference]{IEEEtran}

\IEEEoverridecommandlockouts
\usepackage{balance}
\usepackage{amsmath}
\usepackage{amssymb}
\usepackage{multicol}

\usepackage{hyperref}
\usepackage{ifpdf}

   \usepackage[pdftex]{graphicx}
   \graphicspath{{./pdf/}}
   \DeclareGraphicsExtensions{{.pdf}}

\usepackage[usenames]{color}

\begin{document}
\hyphenation{multi-symbol}
\title{A New Analysis of the DS-CDMA \\ Cellular Downlink Under Spatial Constraints}
\author{\IEEEauthorblockN{Matthew C. Valenti,\IEEEauthorrefmark{1}
Don Torrieri,\IEEEauthorrefmark{2}
and Salvatore Talarico\IEEEauthorrefmark{1} }
\IEEEauthorblockA{\IEEEauthorrefmark{1}West Virginia University, Morgantown, WV, USA. \\
\IEEEauthorrefmark{2}U.S. Army Research Laboratory, Adelphi, MD, USA.}
\thanks{M.C. Valenti and S. Talarico were sponsored by the National Science Foundation under Award No. CNS-0750821 and by the United States Army Research Laboratory under Contract W911NF-10-0109.}
}
\date{}
\maketitle

\vspace{-1cm}
\thispagestyle{empty}

\begin{abstract}
The direct-sequence code-division multiple access (DS-CDMA) cellular downlink is modeled by a constrained random spatial model involving a fixed number of base stations placed over a finite area with a minimum separation.  The analysis is driven by a new closed-form expression for the conditional outage probability at each mobile, where the conditioning is with respect to the network realization.  The analysis features a flexible channel model, accounting for path loss, Nakagami fading, and shadowing.    By generating many random networks and applying a given resource allocation policy, the distribution of the rates provided to each user is obtained.  In addition to determining the average rate, the analysis can determine the transmission capacity of the network and can characterize fairness in terms of the fraction of users that achieve a specified rate.  The analysis is used to compare a rate-control policy against a power-control policy and investigate the influence of the minimum base-station separation.
\end{abstract}

\section{Introduction} \label{Section:Intro}
The classical approach to analyzing the cellular downlink is to model the network as a lattice or regular grid of base stations \cite{lee:1986}.  By using the geometry of the grid, which typically assumes a honeycomb-like arrangement of hexagons, along with models for the underlying propagation conditions and channel reuse strategies, performance metrics can be computed at various potential mobile locations.  Often, the analysis focuses on the worst-case locations, which are at the cell edges.  The hexagonal grid model was used to analyze the other-cell interference (OCI) of a power-controlled direct-sequence code-division multiple access (DS-CDMA) downlink in \cite{viterbi:1994}.  Although conceptually simple and locally tractable, the grid assumption is a poor model for actual base-station deployments, which cannot assume a regular grid structure due to a variety of regulatory and physical constraints.

Recently, a new approach to the analysis of cellular networks has been proposed in \cite{andrews:2011} and \cite{dec:2010}, which models the base-station locations as a realization of a random point process, thereby allowing the use of analytical tools from stochastic geometry \cite{stoyan:1996}. This approach can be used to determine the performance of a typical mobile user.  By combining the effects of random base-station location, fading, and shadowing into a single random variable, performance metrics such as coverage probability (the complement of the outage probability) and average achievable rate can be determined. Under certain limitations, the performance metrics can be found in closed form with surprisingly simple expressions, and these expressions can provide insight into the influence of key network and channel parameters such as the path-loss exponent, the density of base stations, and the minimum required signal-to-interference-and-noise ratio (SINR) required to achieve acceptable coverage.   More recently, this approach has been extended to account for multi-tier heterogenous networks \cite{dhillon:2012}.

The grid-based and random-spatial approaches represent two extreme perspectives in modeling cellular networks.  Whereas the grid-based approach overly constrains the base-station locations, a pure random-spatial approach is similarly unrealistic because it does not place enough constraints on the base-station placements.  For instance, the spatial model in \cite{andrews:2011} assumes that the base stations are drawn from a two-dimensional Poisson point process (PPP) and that the network extends infinitely on the Euclidian plane.   This is a poor model for several reasons.  The first is that no network has an infinite area.  The second is that any realization of a PPP could have a large, even infinite, number of base stations placed within a finite area.  Finally, the pure PPP model does not permit a minimum separation between base stations, which is characteristic of actual macro-cellular deployments.  


In a typical modern cellular system, each base station is allocated a maximum total transmit power, which must be shared among the users in its cell through some resource allocation policy.  Simple spatial models do not adequately capture the nuances of power allocation policies.    For instance, even a simple equal-share power allocation policy  is difficult to model analytically because the power allocated to a given user will depend on the number of mobiles in the same cell.    As a complement to power control, systems often use rate control to ensure that the rate provided to each user is maximized under a constraint on outage probability.  While average achievable rate is considered in \cite{andrews:2011}, it is computed under the assumption that the rate provided to each mobile perfectly adapts to the instantaneous SINR such that the outage is zero.  This overly optimistic assumption is not realistic for current rate-control implementations, which adapt the modulation and coding scheme to provide a particular outage probability (typically 0.1) when averaged over the fading, but conditioned on the network realization.

In this paper, we use a constrained random spatial approach to model the DS-CDMA downlink.  The spatial model places a fixed number of base stations within a region of finite extent.  The model enforces a minimum separation among the base stations.  The model for base-station placement is a binomial point process (BPP) with repulsion,   which we call a {\em uniform clustering} model.  To facilitate the study of resource allocation policies that depend on the mobile locations, the mobiles are also placed according to a uniform clustering process with a higher density and smaller minimum separation than that used to place the base stations.

The analysis in this paper is driven by a new closed-form expression, recently published in \cite{torrieri:2012}, for the {\em conditional} outage probability at each mobile, where the conditioning is with respect to the network realization.  The approach involves drawing realizations of the network according to the desired spatial and shadowing models, and then computing the outage probability at each realized mobile location.  Because the outage probability at each mobile is averaged over the fading, it can be found in closed form with no need to simulate the corresponding channels.  A Nakagami-m fading model is assumed, which models a wide class of channels, and the Nakagami-m fading parameters do not need to be identical for all communication links.  This is a useful feature that can be used to model situations where the base station serving a mobile is in the line of sight (LOS) while the interfering base stations are non-LOS.

By averaging over many network realizations, the mean outage probability can be found for a typical mobile.  However, characterizing only the average performance is of limited utility, and the approach presented in this paper allows the analysis to extend beyond the ergodic performance of the typical mobile user.  For instance, by averaging over the mobiles that are farthest from their base station, the performance of cell-edge mobiles can be determined.  More generally, the outage probability of each mobile can be constrained, and the statistics of the rate provided to each user can be determined under various power-control and rate-control policies.  By plotting the complementary cumulative distribution function (ccdf) of the per-user rate, the variability in mobile performance can be visualized.

The remainder of this paper is organized as follows. Section \ref{Section:SystemModel} presents a model of the network culminating in an expression for SINR.  Section \ref{Section:Outage} provides an expression for the outage probability based on the analysis published in \cite{torrieri:2012}.  Section \ref{Section:Policies} discusses policies for rate control and power control.  A performance analysis is given in Section \ref{Section:Performance}, which compares power control with rate control on the basis of average rate, transmission capacity, and fairness.  The Section also investigates the influence of the  minimum base-station separation.   Finally, the paper concludes in Section \ref{Section:Conclusion}.


\section{Network Model} \label{Section:SystemModel}
\begin{figure}[t]
\centering
\hspace{-0.25cm}
\includegraphics[width=8.25cm]{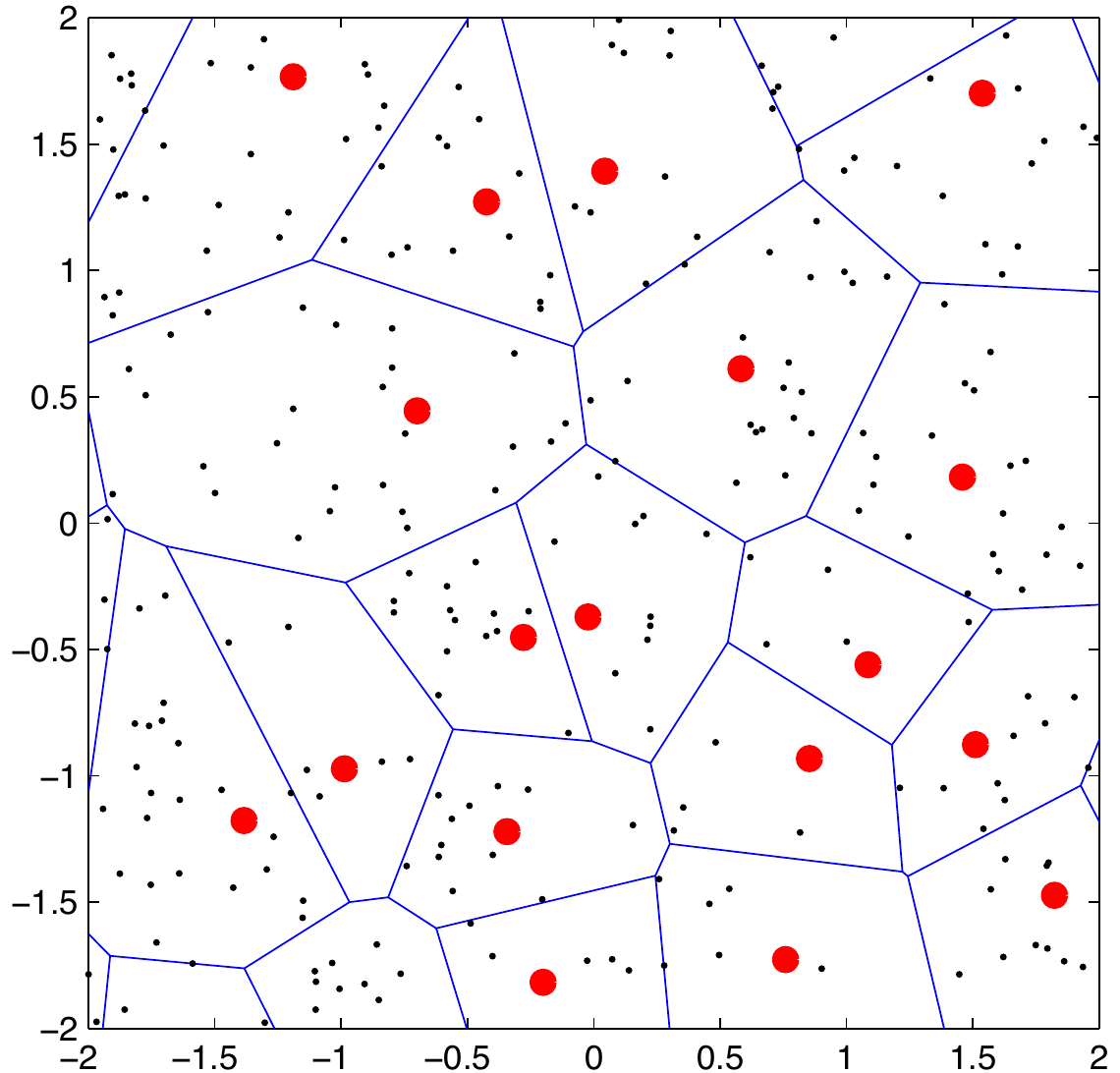}
\vspace{-0.25cm}
\caption{ Close-up of an example network topology.  Base stations are represented by large filled circles, and mobiles by small dots. Cell boundaries are indicated, and the minimum base-station separation is $r_{bs} = 0.25$.  The average cell load is $K/M = 16$ mobiles.
\label{Figure:FigA} }
\vspace{-0.6cm}
\end{figure}

The network comprises $M$ cellular base stations $\{X_1, ..., X_M\}$ and $K$ mobiles $\{ Y_1, ..., Y_K\}$ placed on a disk of radius $r_{net}$ and area $A_{net} = \pi r^2_{net}$.  The variable $X_i$ represents both the $i^{th}$ base station and its location, and similarly, $Y_j$ represents the $j^{th}$ mobile and its location.  An {\em exclusion zone} of radius $r_{bs}$ surrounds each base station, and no other base stations are allowed within this zone.  Similarly, an exclusion zone of radius $r_{m}$ surrounds each mobile, and no other mobiles are allowed within a placed mobile's exclusion zone.  Fig. \ref{Figure:FigA} shows a portion of an example network with average number of mobiles per cell $K/M = 16$, a base-station exclusion radius $r_{bs} = 0.25$, and a mobile exclusion radius $r_m = 0.01$.  The base station locations are given by the large filled circles, and the mobiles are dots.  A Voronoi tessellation shows the cell boundaries that occur in the absence of shadowing.

Each mobile connects to at most one base station.
Let $\mathcal Y_i$ be the set of mobiles connected to base station $X_i$,
and $K_i = | \mathcal Y_i |$ be the number of mobiles served by $X_i$.  Furthermore, let $\mathsf{g}(j)$ be a function that returns the index of the base station serving $Y_j$ so that $Y_j \in \mathcal Y_i$ if $\mathsf{g}(j)=i$.  If $Y_j$ cannot connect to any base station, which is possible when a cell runs out of available channels, then $\mathsf{g}(j) = 0$.

The downlink signals use orthogonal DS-CDMA sequences with common spreading factor $G$.  Because the sequences transmitted by a particular base station are orthogonal, the only source of intracell interference is due to multipath and the corresponding loss of orthogonality.  However, if the ratio of the maximum-power and minimum-power multipath components is sufficiently small (e.g., less than about $0.1G$), then the multipath components will have negligible effect.  For this reason, we neglect the intracell interference and assume that intercell interference is the only source of interference.

The signal is transmitted by base station $X_i$ to mobile $Y_{j}$ with average power $P_{i,j}$.  We assume that the base stations transmit with a common power $P_0$ such that
\begin{eqnarray}
\frac{1}{1-f_p}
\sum_{j: Y_j \in \mathcal Y_i}
P_{i,j}
& = &
P_0
\label{eqn:pwr_constraint}
\end{eqnarray}
for each $i$, where $f_{p}$ is the fraction of the base-station power reserved for pilot signals needed for synchronization and channel estimation.  Power allocation strategies are considered later in this paper.

Using spreading sequences with a spreading factor of $G$ directly reduces the power of the intercell interference. While the intracell sequences transmitted by a particular cell's base station are synchronous, the varying propagation delays from the other base stations cause the intercell interference to be asynchronous.  Because of this asynchronism, the intercell interference is further reduced by the chip factor $h(\tau_{i,k})$, which is a function of the chip waveform and the timing offset $\tau_{i,k}$ at the mobile between the signal received from interfering base station $X_k$ and the signal received from serving base station $X_i$.   
When the timing offset $\tau_{i,k}$ is assumed to have a uniform distribution over the chip interval, 
then the expected value of $h(\tau_{i,k})$ is 2/3 \cite{torrieri:2011}. It is assumed henceforth that $G/h(\tau_{i,k})$ is a
constant equal to $G/h$ at each mobile in the network.


The power of $X_i$ received at $Y_j$ also depends on the fading and path loss models.  We assume that path loss has a power-law dependence on distance and is perturbed by shadowing.  When accounting for fading and path loss, the despread instantaneous power of $X_i$ at mobile $Y_j$ is
\begin{eqnarray}
\rho_{i,j}
& = &
\begin{cases}
{P}_{i,j} g_{i,j} 10^{\xi_{i,j}/10} f\left(  ||X_{i}-Y_{j}||\right) & \mbox{if $\mathsf{g}(j) = i$} \vspace{0.2cm} \\
\left( \frac{h}{G} \right) {P}_{i,j} g_{i,j} 10^{\xi_{i,j}/10} f\left(  ||X_{i}-Y_{j}||\right) & \mbox{if $\mathsf{g}(j) \neq i$}
\end{cases} \nonumber \\
\label{eqn:power}%
\end{eqnarray}
where $g_{i,j}$ is the power gain due to fading, $\xi_{i,j}$ is a
\textit{shadowing factor}, and $f(\cdot)$ is a path-loss function. The
\{$g_{i,j}\}$ are independent with unit-mean, and $g_{i,j}=a_{i,j}^{2}$,
where $a_{i,j}$ is Nakagami with parameter $m_{i,j}$.
While the $\{g_{i,j}\}$ are independent from mobile to mobile, they are not
necessarily identically distributed, and each mobile can have a distinct Nakagami parameter $m_{i,j}$.
When the channel between $X_i$ and $Y_j$ experiences Rayeigh fading, $m_{i,j}=1$
and $g_{i,j}$ is exponentially distributed.
In the presence of log-normal shadowing, the $\{\xi_{i,j}\}$ are
i.i.d. zero-mean Gaussian with variance $\sigma_{s}^{2}$.
In the absence of shadowing, $\xi_{i,j}=0$. \ For $d \geq d_{0}$, the path-loss function is
expressed as the attenuation power law
\begin{equation}
f\left(  d\right)  =\left(  \frac{d}{d_{0}}\right)  ^{-\alpha}
\label{eqn:pathloss}%
\end{equation}
where $\alpha\geq2$ is the attenuation power-law exponent, and
that $d_{0}$ is sufficiently large that the signals are in the far field.

The base station $X_{\mathsf{g}(j)}$ that serves mobile $Y_j$ is selected to be the
one with index
\begin{eqnarray}
\mathsf{g}
(j)
& = &
\underset{i}{\operatorname{argmax}}
\,
\left\{
10^{\xi_{i,j}/10} f\left(  ||X_{i}-Y_{j}||\right)
\right\}
\label{eqn:connectivity}
\end{eqnarray}
which is the base station with minimum path loss to $Y_j$.  In the absence of shadowing, it will be the base station that is closest to $Y_j$.  In the presence of shadowing, a mobile may actually be associated with a base station that is more distant than the closest one, if the shadowing conditions are sufficiently better.  We assume a maximum of $G$ orthogonal spreading sequences per cell, and once $G$ users are connected to a base station, no more can be served.


The instantaneous SINR at mobile $Y_j$ is
\begin{eqnarray}
\gamma_j
& = &
\frac{\rho_{\mathsf{g}(j),j}}{\displaystyle{\mathcal{N}}
+
\mathop{ \sum_{i=1}^M }_{ i \neq \mathsf{g}(j) }
\rho_{i,j}}
\label{eqn:SINR1}
\end{eqnarray}
where $\mathcal{N}$ is the noise power. Substituting
(\ref{eqn:power}) and (\ref{eqn:pathloss}) into (\ref{eqn:SINR1}) yields
\vspace{-0.3cm}
\begin{eqnarray}
\gamma_j
&  = &
\frac{g_{\mathsf{g}(j),j}\Omega_{\mathsf{g}(j),j}}
{\displaystyle\Gamma^{-1}
+
\frac{h}{G}
\mathop{ \sum_{i=1}^M }_{ i \neq \mathsf{g}(j) }
g_{i,j}\Omega_{i,j}}
\label{Equation:SINR2}
\end{eqnarray}
where $\Gamma=d_{0}^{\alpha}P_{0}/\mathcal{N}$ is the signal-to-noise ratio
(SNR) at a mobile located at unit distance when fading and
shadowing are absent, 
and
\vspace{-0.2cm}
\begin{eqnarray}
\Omega_{i,j}
& = &
\frac{ P_{i,j}}{P_0} 10^{\xi_{i,j}/10} ||X_i-Y_j||^{-\alpha}
\label{eqn:omega}%
\end{eqnarray}
is the normalized power of $X_i$ at receiver $Y_j$ before despreading.

\section{Outage Probability} \label{Section:Outage}
\label{Section:OutageProbability}
Let $\beta_j$ denote the minimum SINR required by $Y_j$ for reliable reception and $\boldsymbol{\Omega }_j=\{\Omega_{1,j},...,\Omega _{M,j}\}$ represent the set of normalized despread base-station powers received by $Y_j$.  An \emph{outage} occurs when the SINR falls below $\beta_j$.  As discussed subsequently, there is a relationship between the SINR threshold and the supported {\em rate} of the transmission.  Conditioning on $\boldsymbol{\Omega }_j$, the outage probability of mobile $Y_j$ is
\begin{eqnarray}
   \epsilon_j
   & = &
   P \left[ \gamma_j \leq \beta_j \big| \boldsymbol \Omega_j \right].
   \label{Equation:Outage1}
\end{eqnarray}
Because it is conditioned on $\boldsymbol{\Omega }_j$, the outage probability depends on the particular network realization, which has dynamics over timescales that are much slower than the fading.
By defining a variable
\vspace{-0.25cm}
\begin{eqnarray}
  \mathsf Z_j & = & \beta_j^{-1} g_{\mathsf{g}(j),j} \Omega_{\mathsf{g}(j),j}
  -
  \frac{h}{G}\mathop{ \sum_{i=1}^M }_{ i \neq \mathsf{g}(j) }
  \Omega_{i,j} \label{eqn:z}
\end{eqnarray}
the conditional outage probability may be expressed as
\begin{eqnarray}
  \epsilon_j
  & = &
  P
  \left[
   \mathsf Z_j  \leq \Gamma^{-1} \big| \boldsymbol \Omega_j
  \right]
  = F_{\mathsf Z_j} \left( \Gamma^{-1} \big| \boldsymbol \Omega_j \right) \label{Equation:OutageCDF}
\end{eqnarray}
which is the cumulative distribution function (cdf) of $\mathsf Z_j$ conditioned on $\boldsymbol \Omega_j$ and evaluated at $\Gamma^{-1}$.

Define $\bar{F}_{\mathsf{Z}_j}(z | \boldsymbol \Omega_j) = 1 - F_{\mathsf{Z}_j}(z | \boldsymbol \Omega_j)$ to be the
complementary cdf of $\mathsf{Z}_j$ conditioned on $\boldsymbol \Omega_j$. Restricting the Nakagami parameter $m_{\mathsf{g}(j),j}$ between mobile $Y_j$ and its serving base station $X_{\mathsf{g}(j)}$ to be integer-valued, the complementary
cdf of $\mathsf{Z}_j$ conditioned on $\boldsymbol{\Omega}_j$ is proved in \cite{torrieri:2012} to be
\begin{eqnarray}
\bar{F}_{\mathsf Z_j}\left( z \big| \boldsymbol \Omega_j \right)
& = &
e^{-\beta_0 z }
\sum_{n=0}^{m_0-1} {\left( \beta_0 z \right)}^n
 \sum_{k=0}^n
\frac{ z^{-k} H_k ( \boldsymbol \Psi )}{ (n-k)! }
\label{Equation:NakagamiConditional}
\end{eqnarray}
where $m_0 = m_{\mathsf{g}(j),j}$, $\beta_0 = \beta m_0/\Omega_0$,
\begin{eqnarray}
   \Psi_i
   & = &
   \left(
       \frac{\beta_0 h \Omega_{i,j}}{G m_{i,j} } + 1
    \right)^{-1}\hspace{-0.5cm}    \label{Equation:Psi}\\
   H_k ( \boldsymbol \Psi )
   & = &
   \mathop{ \sum_{\ell_i \geq 0}}_{\sum_{i=0}^{M}\ell_i=k}
   \left(
   \mathop{ \prod_{i=1}^M }_{ i \neq \mathsf{g}(j) }
   G_{\ell_i} ( \Psi_i )
   \right), \label{Equation:Hfunc}
\end{eqnarray}
the summation in (\ref{Equation:Hfunc}) is over all sets of indices that sum to $k$, and
\begin{eqnarray}
 G_\ell( \Psi_i )
 & = &
 \frac{\Gamma( \ell + m_{i,j} ) }
 {\ell! \Gamma( m_{i,j} ) }
 \left( \frac{\Omega_{i,j}}{ m_{i,j} } \right)^{\ell} \Psi_i^{ m_{i,j} +\ell}.
 \label{Equation:Gfunc}
\end{eqnarray}

\section{Policies}\label{Section:Policies}
A key consideration in the operation of the network is the manner that the total power $P_0$ transmitted by a base station is shared by the mobiles it serves, which influences the rates provided to each user.   This section discusses two options for allocating rate and power, {\em rate control} and {\em power control}.  

\subsection{Rate Control}

A simple and efficient way to allocate $P_0$ is with an {\em equal-share} policy, which involves base station $X_i, i=\mathsf{g}(j)$, transmitting to mobile $Y_j$ with power
\begin{eqnarray}
  P_{i,j}
  & = &
  \frac{P_0}{K_i(1-f_p)}. \label{Equation:Share}
\end{eqnarray}
Under this policy, the SINR will vary dramatically at each mobile.  If a common SINR threshold $\beta_j$ is used by all mobiles, then the outage probability will likewise be highly variable.    Instead of using a fixed threshold, the threshold $\beta_j$ of mobile $Y_j$ can be selected such that the outage probability of mobile $Y_j$ satisfies the constraint $\epsilon_j = \hat{\epsilon}$.  A constraint of $\hat{\epsilon}=0.1$ is typical and appropriate for modern systems that use a hybrid automatic repeat request (HARQ) protocol.


For a given $\beta_j$, there is a corresponding transmission rate $R_j$ that can be supported.   Let $R_j = C(\beta_j)$ represent the relationship between $R_j$, expressed in units of bits per channel use (bpcu), and $\beta_j$.  The relationship depends on the modulation and coding schemes used, and typically only a discrete set of $R_j$ can be supported.   While the exact dependence of $R_j$ on $\beta_j$ can be determined empirically through tests or simulation,  we make the simplifying assumption when computing our numerical results that $C(\beta_j) = \log_2(1+\beta_j)$ corresponding to the Shannon capacity.  This assumption is fairly accurate for capacity-approaching codes with an infinite number of possible rates.  This is a reasonable model for modern cellular systems,
which use turbo codes with a large number of available rates.

With an equal power share, the number of mobiles $K_i$ in the cell is first determined, and then the power share given to each is found from (\ref{Equation:Share}).  For each mobile in the cell, the corresponding $\beta_j$ that achieves the outage constraint $\epsilon_j = \hat{\epsilon}$ is found by inverting the outage probability expression given in Section \ref{Section:Outage}.  Once $\beta_j$ is found, the corresponding $R_j$ is found by using the function $R_j = C(\beta_j)$.  The rate $R_j$ is adapted by changing the number of information bits per channel symbol. The processing gain $G$ and symbol rate are held constant, so there is no change in bandwidth.  Because this policy involves fixing the transmit power and then determining the rate for each mobile that satisfies the outage constraint, we refer to it as {\em rate control}.

\subsection{Power Control}

A major drawback with rate control is that the rates provided to the different users in the network will vary significantly.  This variation will result in unfairness to some users, particularly those located at the edges of the cells.  To ensure fairness, $R_j$ could be constrained to be the same for all $X_j \in \mathcal Y_i$.  The power transmitted to mobile $Y_j$ is then found such that the mobile has outage probability $\epsilon_j = \hat{\epsilon}$, while the total power provided to all mobiles in the cell satisfies power constraint (\ref{eqn:pwr_constraint}).  Because the power transmitted to each mobile is varied while holding the rate constant for all mobiles in the cell, we refer to this policy as {\em power control}.  Note that, while the rate is the same for all users within a given cell, it may vary from cell to cell.  In particular, the rate of a given cell is found by determining the value of $R$ that allows the outage constraint $\epsilon_j = \hat{\epsilon}$ and power constraint (\ref{eqn:pwr_constraint}) to be simultaneously met.

%

\section{Performance Analysis}\label{Section:Performance}
Under outage constraint $\hat{\epsilon}$, the performance of a given network realization is largely determined by the set of achieved rates $\{R_j\}$ of the $K$ users in the network.  Because the network realization is random, it follows that the set of rates is also random.  Let the random variable $R$ represent the rate of an arbitrary user.  The statistics of $R$ can be found for a given class of networks using a Monte Carlo approach as follows.  Draw a realization of the network by placing $M$ base stations and $K$ mobiles within the disk of radius $r_{net}$ according to the uniform clustering model with minimum base-station separation $r_{bs}$ and minimum mobile separation $r_m$.  Compute the path loss from each base station to each mobile, applying randomly generated shadowing factors if shadowing is present.  Determine the set of mobiles associated with each base station.
At each base station, apply the power allocation policy to determine the power it transmits to each mobile that it serves.  By setting the outage equal to the outage constraint, invert the outage probability expression to determine the SINR threshold for each mobile in the cell.  By applying the function $R_j=C(\gamma_j)$, find the rate of the mobile.  Repeat this process for a large number of networks, all with the same spatial constraints.

Let $E[R]$ represent the mean value of the variable $R$, which can be found by numerically averaging the values of $R$ obtained using the procedure described in the previous paragraph.  While $E[R]$ is a useful metric, it does not account for the loss in throughput due to the inability to successfully decode during an outage, and it does not account for the spatial density of transmissions.   These considerations are taken into account by the \emph{transmission capacity}, defined as \cite{weber:2010}
\begin{eqnarray}
  \tau
  & = &
  \lambda
  \left( 1 - \hat{\epsilon} \right)
  E[R]
  \label{eqn:tc}
\end{eqnarray}
where $\lambda = K/A_{net}$ is the density of transmissions in the network.  Transmission capacity can be interpreted as the spatial intensity of transmissions; i.e., the rate of successful data transmission per unit area.

As an example, consider a network with $M=50$ base stations placed in a network of radius $r_{net} = 2$ with base-station exclusion zones of radius $r_{bs} = 0.25$.  A variable number $K$ of mobiles are placed within the network using exclusion zones of radius $r_m = 0.01$.  The outage constraint is set to $\bar{\epsilon} = 0.1$ and both power control and rate control are considered. A mobile in an overloaded cell is denied service.  In particular, the $K_i - G$ mobiles whose path losses from the base station are greatest are denied service, in which case they do not appear in the set $\mathcal Y_i$ for any $i$, and their rates are set to $R_j = 0$.   The SNR is set to $\Gamma = 10$ dB, the fraction of power devoted to pilots is $f_p = 0.1$, and the spreading factor set to $G=16$ with chip factor $h=2/3$.  The propagation environment is characterized by a path-loss exponent $\alpha = 3$, and the Nakagami factors are $m_{i,j} = 3$ for $i = \mathsf{g}(j)$ while $m_{i,j} = 1$ for $i \neq \mathsf{g}(j)$; i.e., the signal from the serving base station experiences milder fading than the signals from the interfering base stations.  This is a realistic model because the signal from the serving cell is likely to be in the line of sight (LOS), while the interfering cells are typically not LOS.

\begin{figure}[t]
\centering
\includegraphics[width=8.75cm]{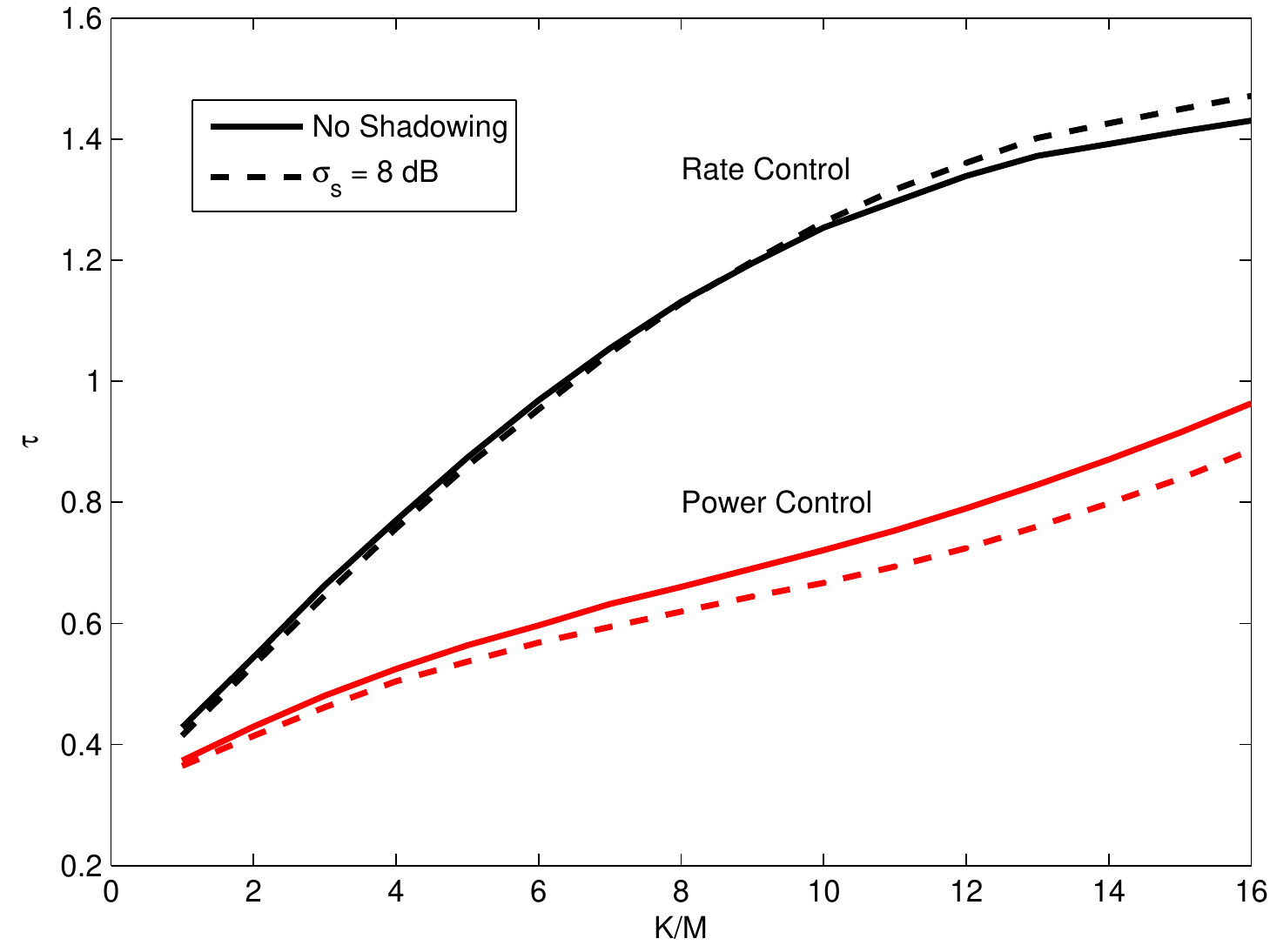}
\vspace{-0.55cm}
\caption{ Transmission capacity of the Example as a function of $K/M$ with rate control and power control.
\label{Figure:FigB} }
\vspace{-0.65cm}
\end{figure}

Fig. \ref{Figure:FigB} shows the transmission capacity, as a function of the ratio $K/M$, of rate control and power control in an unshadowed environment as well as in the presence of shadowing with $\sigma_s = 8$ dB.  The figure shows that the transmission capacity under rate control is higher than it is under power control.  This disparity occurs because mobiles that are close to the base station are able to be allocated extremely high rates under rate control, while with power control, mobiles close to the base station must be allocated the same rate as mobiles at the edge of the cell.  As the network becomes denser ($K/M$ increases), shadowing actually improves the performance with rate control, while it degrades the performance with power control.  This is because shadowing can sometimes cause the signal power of the base station serving a mobile to increase, while the powers of the interferers are reduced.  The effect of favorable shadowing is equivalent to the mobile being located closer to its serving base station. When this occurs with rate control, the rate is increased, sometimes by a very large amount.  While shadowing does not induce extremely favorable conditions very often,  when it does the improvement in rate is significant enough to cause the average to increase.
On the other hand, a single user with favorable shadowing conditions operating under power control will not improve the average rate because all mobile in a cell receive the same rate.

\begin{figure}[t]
\centering
\includegraphics[width=8.75cm]{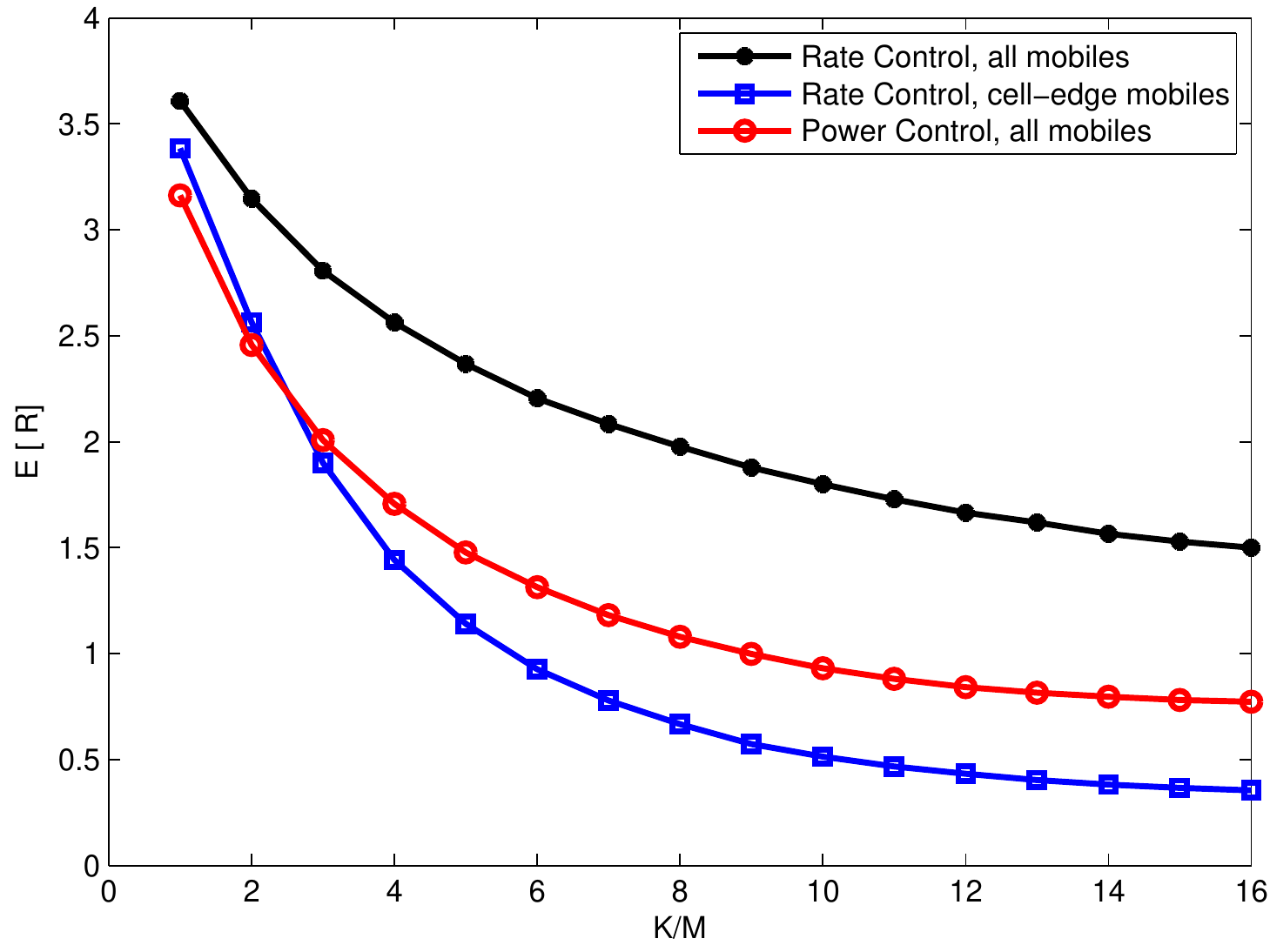}
\vspace{-0.55cm}
\caption{ Average rate of the Example  as a function of $K/M$ in the presence of shadowing with rate control and power control.  For rate control, the averaging is done over all mobiles and over just the cell-edge mobiles.  With power control, all mobiles in a cell are given the same rate. \label{Figure:FigC} }
\vspace{-0.5cm}
\end{figure}

While rate control offers a higher {\em average} rate than power control, the rates it offers are much more variable.  This behavior can be seen in Fig. \ref{Figure:FigC}, which compares the rates of all users against those located at the cell edges.  In particular, the figure shows the rate averaged across all mobiles for both power control and rate control, as well as the rate averaged across just the cell-edge mobiles for rate control, where the cell-edge mobiles are defined to be the 5 percent of the mobiles that are furthest from their serving base station.  The average rate of cell-edge mobiles is not shown for power control because each cell-edge mobile has the same rate as all the mobiles in the same cell.  As seen in the figure, the performance of cell-edge mobiles is worse with rate control than it is with power control.

\begin{figure}[t]
\centering
\includegraphics[width=8.75cm]{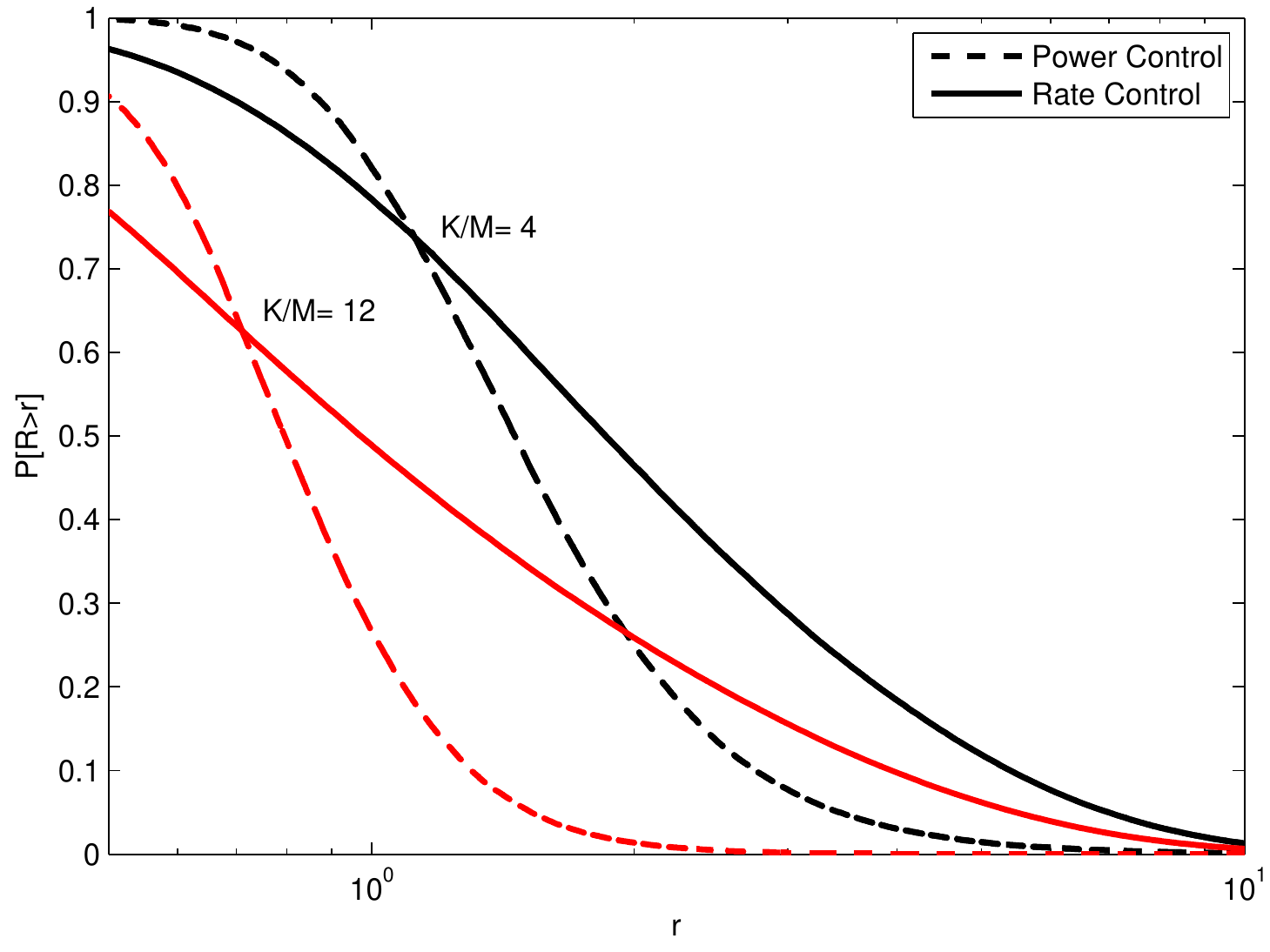}
\vspace{-0.55cm}
\caption{ Ccdf of $R$ with either rate control or power control for a lightly loaded system $(K/M=4)$ and a moderately-loaded system $(K/M=12)$. \label{Figure:FigD}  }
\end{figure}

The fairness of a particular power allocation policy can be further visualized by plotting the ccdf of $R$, which is the probability that $R$ exceeds a threshold $r$; i.e. $P[R>r]$.  Fig. \ref{Figure:FigD} shows the ccdf of $R$ for the Example with shadowing ($\sigma_s = 8$ dB) and either rate control or power control.  Two system loads are considered: a lightly loaded system $(K/M=4)$ and a moderately-loaded system $(K/M=12)$.  The ccdf curves for power control are steeper than they are for rate control, indicating less variability in the provided rates.  The lower variability in rate corresponds to improved fairness.  For instant, with a load $K/M = 4$, almost all (99.9\%) users are provided with rates of at least $r=0.5$ under power control, while with rate control 96\% of the users are provided with rates of at least $r=0.5$ implying that a significant fraction of the users are provided with lower rates.

\balance

\begin{figure}[t]
\centering
\hspace{-0.5cm}
\includegraphics[width=8.75cm]{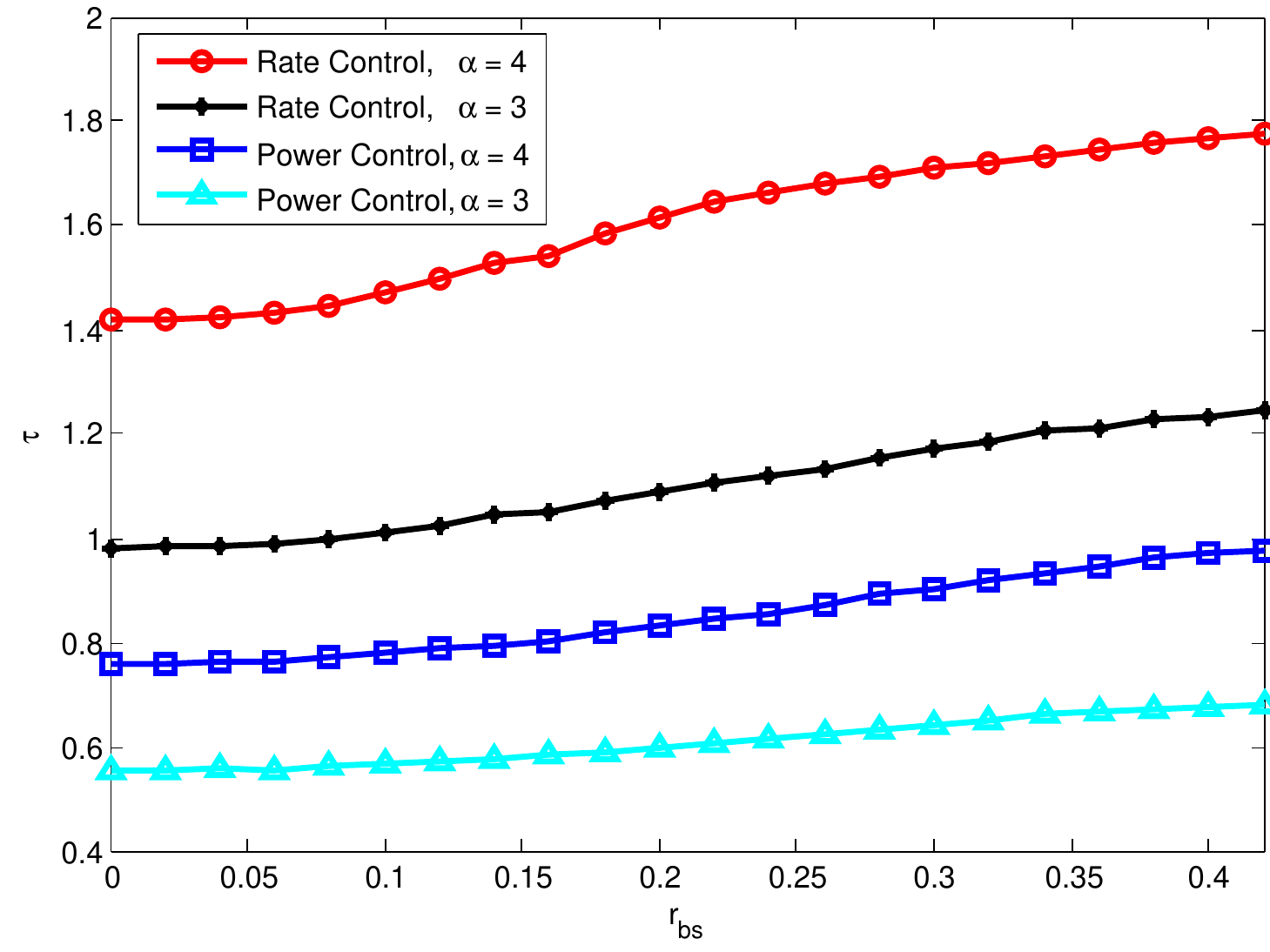}
\caption{ Transmission capacity for rate and power control as function of $r_{bs}$ for $K/M=8$ in Mixed Fading and Shadowing ($\sigma_s$ = 8 dB). \label{Figure:FigF}  }
\end{figure}
A key feature of the proposed analysis is that it permits a minimum spacing of $r_{bs}$ around each base station.  The dependence of the transmission capacity on $r_{bs}$ is shown in Fig. \ref{Figure:FigF} for both rate control and power control with a system load of $K/M = 8$.  Curves are shown for two values of path-loss exponent: $\alpha = 3$ and $\alpha = 4$.  Except for the values of $r_{bs}$ and $\alpha$, the parameters are the same as in the Example with shadowing.   Increasing $r_{bs}$ improves the transmission capacity.  The improvement is slightly more pronounced for rate control than for power control, and is more pronounced for the higher value of $\alpha$.

%
%

\section{Conclusion} \label{Section:Conclusion}
This paper has presented a powerful new approach for modeling and analyzing the cellular downlink.  Unlike simple spatial models, the model in this paper allows constraints to be placed on the distance between base stations, the geographic footprint of the network, and the number of base stations and mobiles.  The analysis features a flexible channel model, accounting for path loss, shadowing, and Nakagami-m fading with non-identical parameters.  The proposed analytical approach provides a way to compare various access and resource allocation techniques.   As a specific application of the model, the paper models a direct-sequence CDMA network, analyzes it using realistic network parameters, and compares the performance of two resource allocation policies.
While this paper only considers resource allocation policies that implement either rate control or power control, other policies that fall between these two extremes could be considered.   While in this paper mobiles in overloaded cells were denied service, the work can be extended to analyze reselection schemes, whereby mobiles in overloaded cells attempt to connect to another nearby base station serving an underloaded cell.  This work could be extended to analyze the uplink and to model other types of access, such as orthogonal frequency-division multiple access (OFDMA).  It could furthermore be extended to handle sectorized cells and coordinated multipoint strategies involving transmissions from multiple base stations.

\bibliographystyle{ieeetr}
\bibliography{gc2012refs}

\end{document}